\documentclass{aa}
\usepackage{psfig,txfonts}

\begin{document}

\title{%The Good, the Bad and the Ugly: 
Unveiling the nature of three 
{\it INTEGRAL} sources \\ through optical spectroscopy\thanks{Based on 
observations collected at the Bologna Astronomical Observatory in Loiano 
(Italy)}}

\author{
N. Masetti,
E. Palazzi,
L. Bassani,
A. Malizia
and J.B. Stephen
}

\institute{
Istituto di Astrofisica Spaziale e Fisica Cosmica --- Sezione di Bologna, 
CNR, via Gobetti 101, I-40129, Bologna (Italy)
}

\offprints{N. Masetti (\texttt{masetti@bo.iasf.cnr.it)}}
\date{Received 10 August 2004; accepted 10 September 2004}

\abstract{
The results of an optical spectroscopy campaign performed at the
Astronomical Observatory of Bologna in Loiano (Italy) on three hard X--ray
sources detected by {\it INTEGRAL} (IGR J17303$-$0601, IGR J18027$-$1455
and IGR J21247+5058) are presented. These data have allowed a
determination of the nature for two of them, with IGR J17303$-$0601 being
a low mass X--ray binary in the Galaxy and IGR J18027$-$1455 a background
Type 1 Seyfert galaxy at redshift $z$ = 0.035. IGR J21247+5058, instead,
has a quite puzzling spectroscopic appearance, with a broad, redshifted
$H_\alpha$ complex superimposed onto a `normal' F/G-type Galactic star
continuum: these features, together with the spatially coincident extended
radio emission, might suggest a chance alignment between a relatively
nearby star and a background radio galaxy. These results underline the 
still non-negligible importance of smaller telescopes in modern 
astrophysics.

\keywords{X--rays: binaries --- X--rays: galaxies --- Techniques: 
spectroscopic --- X--rays: individuals: IGR J17303$-$0601; IGR 
J18027$-$1455; IGR J21247+5058}}

\titlerunning{the nature of three IGR sources}
\authorrunning{N. Masetti et al.}

\maketitle

\section{Introduction}

\begin{figure*}%[t!]
%\begin{center}
\centering{\mbox{\psfig{file=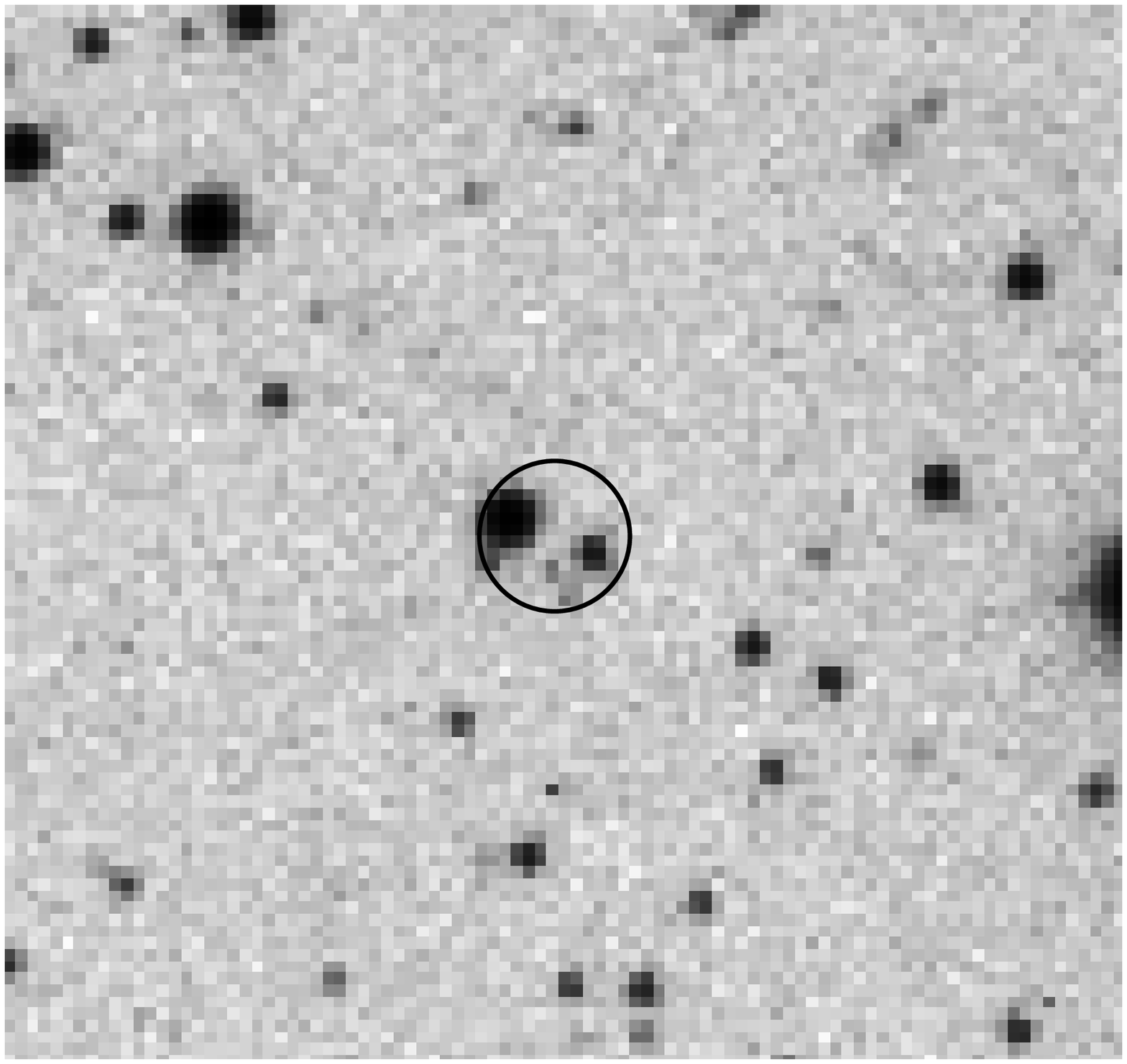,width=5.9cm}}}%\\
%\quad
%\psfig{file=18027.ps,width=6cm}
\centering{\mbox{\psfig{file=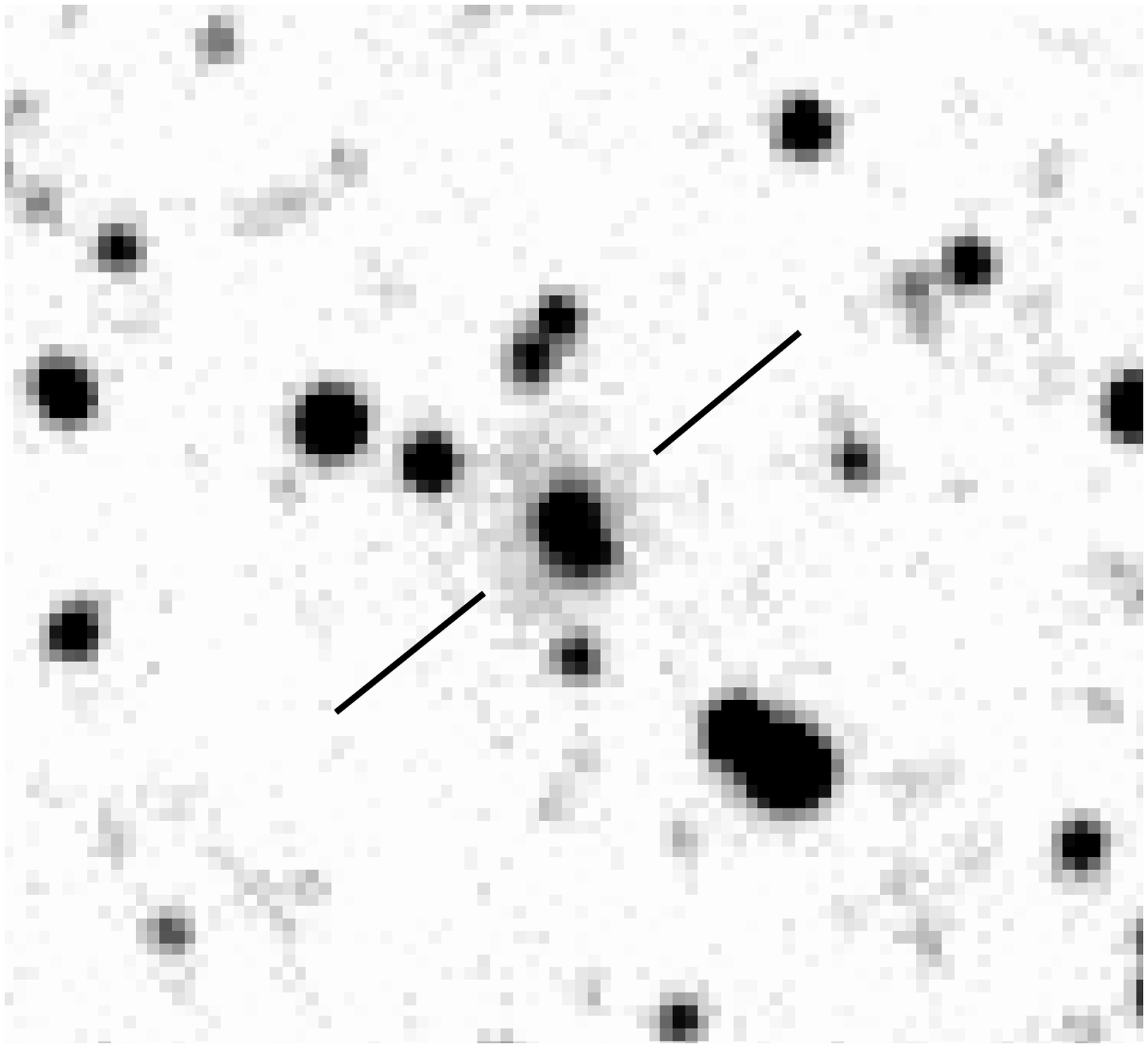,width=5.9cm}}}%\\
\centering{\mbox{\psfig{file=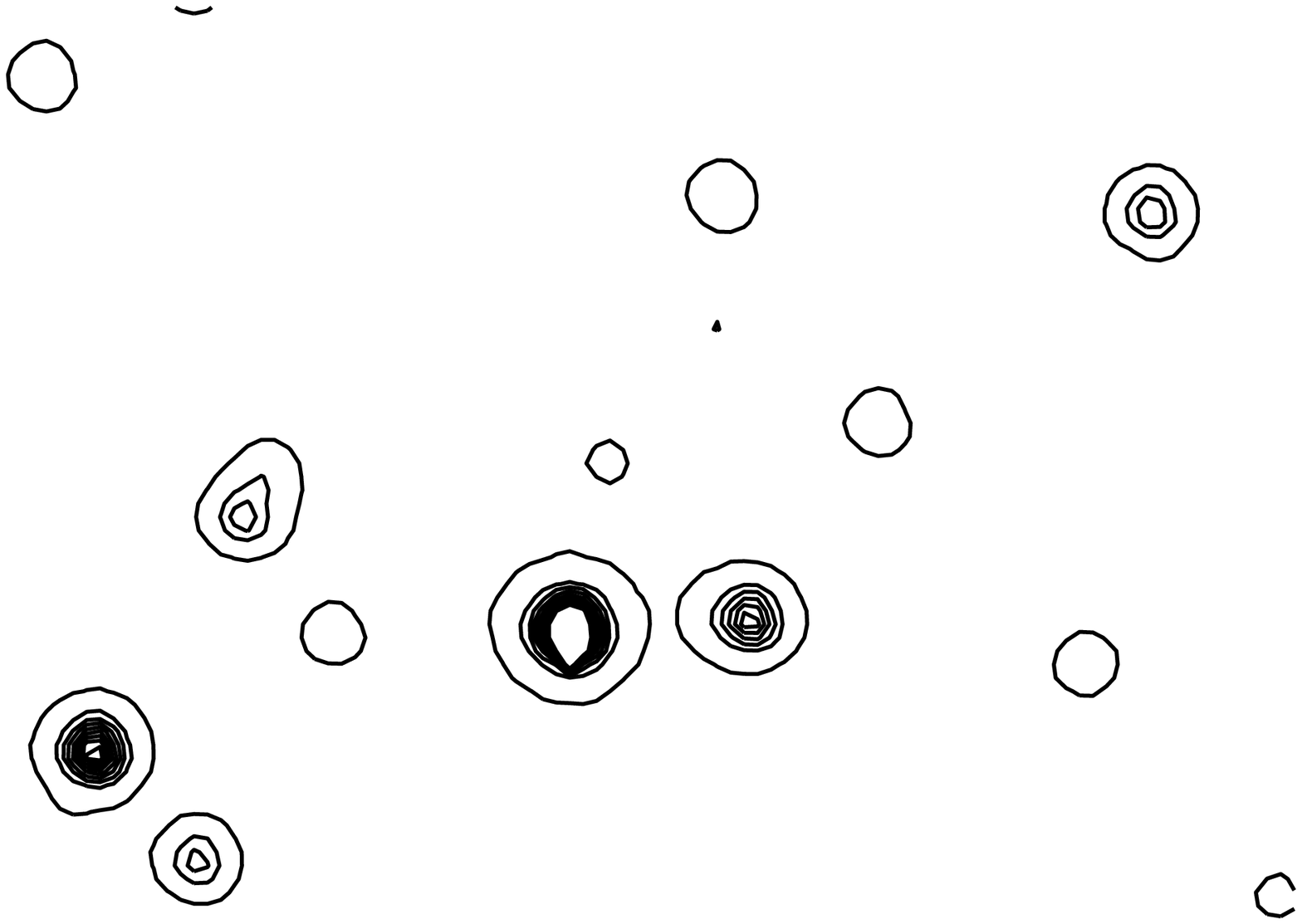,width=5.9cm}}}
%\end{center}
\vspace{-.2cm}
\caption{{\it (Left panel)}: section of an $R$-band image showing the 
{\it ROSAT} error circle (7$''$ radius) of the source 1RXS 
J173021.5$-$055933, positionally consistent with IGR J17303$-$0601.
The field size is about 50$''$$\times$50$''$. {\it (Central 
panel)}: section of an $R$-band image showing the optical counterpart of 
IGR J18027$-$1455 (indicated by the tick marks). The field size is about 
50$''$$\times$50$''$. {\it (Right panel)}: contour plot from the 
$I$-band image of the field around the counterpart of IGR J21247+5058, 
which is the rightmost of the two objects at the centre. 
The field size is about 40$''$$\times$30$''$; the field is rotated of 
about 12$^{\circ}$ clockwise. All images have North up and East to 
the left and were acquired in Loiano on 14-15 July 2004.}
\end{figure*}

\vspace{-.2cm}
One main objective of the {\it INTEGRAL} mission is a regular survey of
the Galactic Plane complemented by a deep exposure of the Galactic Centre.
This makes use of the unique imaging capability of IBIS (Ubertini et al.
2003) which allows the detection of sources at the mCrab level with an
angular resolution of 12$'$ and a typical point source localization
accuracy of 2-3$'$: this has made it possible, for the first time, to 
resolve crowded regions such as the Galactic Centre and the spiral arms, 
and to discover many new hard X--ray objects.

In its first year, IBIS has detected 123 sources between 20 and 100 keV
(Bird et al. 2004): within this sample of hard X-ray emitting objects, 53
low mass and 23 high mass X-ray binaries, 5 Active Galactic Nuclei (AGN)
and a handful of other objects such as pulsars, cataclysmic variables and
a dwarf nova are found. The remaining objects (28, or about 23$\%$ of the
sample) have no obvious conterparts at other wavelengths and therefore
cannot yet be associated with any known class of high-energy emitting
objects. Only for a tiny fraction of these sources have follow-up 
observations at X--ray energies as well as in the optical/near-infrared 
wavebands been carried out so far (Rodriguez 2004).
Although the cross-correlation with catalogues or surveys at other
wavelengths (especially soft X--rays, optical and radio) is of invaluable
help in pinpointing the putative optical candidates, only accurate optical
spectroscopy can confirm the association and reveal the nature of the
object.

Most of these unidentified sources are believed to be X--ray binary 
systems, where one of the two members is either a black hole or a neutron 
star. There is however the possibility that some of them could be AGN 
similar to those already detected (Bassani et al. 2004). 
With the aim of pursuing secure identifications for some of the IBIS
unidentified sources, we have extracted a small sample of three objects
with relatively bright putative optical counterparts in order to assess
their nature. These are IGR J17303$-$0601, IGR J18027$-$1455 and IGR 
J21247+5058.

IGR J17303$-$0601 positionally coincides with 1RXS J173021.5$-$055933
(Voges et al. 1999), an X--ray object detected from 0.1 to 30 keV (by {\it
ROSAT}, {\it HEAO-A1} and {\it RXTE}). Within the small {\it ROSAT} error
box (7$''$) two optical objects, one with $R \sim 15.5$ and the other with
$R \sim 18$ according to the USNO catalogues (Monet et al. 2003), are
found on the Digitized Sky 
Survey\footnote{\texttt{http://archive.eso.org/dss/dss/}} (DSS).
This X--ray source was suggested to be an AGN by Sazonov \& Revnivtsev
(2004) on the basis of its similarity with the spectral slope of this
class of sources as detected by {\it RXTE}.

IGR J18027$-$1455 (Walter et al. 2004) is another likely AGN candidate:
inside the 2$'$ ISGRI error box of this source a {\it ROSAT} object (1RXS
J180245.5$-$145432; Voges et al. 1999) and a
NVSS\footnote{\texttt{http://www.cv.nrao.edu/nvss/}} radio object
(NVSS~180247$-$145451), positionally consistent with each other, are found
(as first noted by Combi et al. 2004a,b). At the radio coordinates, 
an extended
2MASS\footnote{\texttt{http://www.ipac.caltech.edu/2mass/}} infrared
source (2MASXi J1802473$-$145454) is present. The DSS also shows an
extended optical object, with USNO-A2.0 magnitude $R \sim 15$, at this
position. A preliminary analysis of the data which will be presented here,
and indicating that this object is indeed a Type 1 AGN, was reported by
Masetti et al. (2004a).

IGR J21247+5058 (Walter et al. 2004) has recently been associated by
Rib\'o et al. (2004) and Combi et al. (2004b) with the radio source
4C50.55, which shows a morphology typical of a radio galaxy (Mantovani et
al. 1982). The optical counterpart is clearly detected on the DSS and is
present on the USNO-A2.0 catalogue with magnitude $R \sim 15.5$.

To firmly establish the nature of these sources, we thus performed a
spectroscopic campaign on their optical candidates listed above with the
1.5m ``G.D. Cassini" telescope of the Astronomical Observatory of Bologna
located at Loiano (Italy), in order to study the continuum and to reveal
the possible presence of Balmer lines or other features that will pinpoint
the redshift and nature of these objects.

\vspace{-.3cm}
\section{Optical observations at Loiano}

\vspace{-.2cm}
Medium-resolution optical spectra of the optical candidates
of the sources IGR J17303$-$0601, IGR J18027$-$1455 and IGR J21247+5058 
were acquired on 14-15 July 2004 and 2-3 August 2004 in Loiano (Italy) 
with the Bologna Astronomical Observatory $1.52$~metre ``G.D. Cassini'' 
telescope plus BFOSC, equipped with a $1300\times1340$ pixels EEV CCD. 
In all cases Grism \#4 was used, providing a 3500-8700 \AA~nominal 
spectral coverage; slit widths of $2''$ and 2$\farcs$5 during the July and 
August runs, respectively, were used in order to match the night's seeing. 
The use of these setups secured a final dispersion of $4.0$~\AA/pix for 
all spectra.

Besides all the above, in the case of IGR J17303$-$0601 we put the
slit at position angle PA = +65$^{\circ}$ in order to include both
candidate optical counterparts (see Sect. 1); similarly, when we observed
IGR J21247+5058 we rotated the slit by 12$^{\circ}$ to get the spectrum of
a pointlike object lying $\sim$8$''$ southeast of the optical source
coincident with the NVSS radio emission.

Spectra, after correction for flat-field, bias and cosmic-ray
rejection, were background subtracted and optimally extracted (Horne
1986) using IRAF\footnote{\texttt{http://iraf.noao.edu/}}.
Wavelength calibration of the spectra was performed using He-Ar 
lamps, while the flux calibration was applied through the 
spectrophotometric standard BD+25$^{\circ}$3941; finally all spectra 
from each source were stacked together to increase the S/N ratio.
Wavelength calibration was checked by using the positions of background
night sky lines; the error was $\sim$0.5~\AA~for all spectra.

On the night between 14 and 15 July 2004 we also acquired $R$-band 
imaging of IGR J17303$-$0601 and IGR J18027$-$1455, and $BVRI$ imaging
of IGR J21247+5058, again in Loiano with BFOSC under an average 
seeing of 1$\farcs$6.
The EEV CCD, with a scale of 0$\farcs$58/pix, secured a field of
12$\farcm$6$\times$12$\farcm$6. Images were corrected for bias and
flat-field in the usual fashion and calibrated using the PG 2213-006
field (Landolt 1992); the calibration accuracy is better than 2\% in all 
bands. All of the putative counterparts of IGR sources (in Fig. 1) were 
well detected in all images. Magnitudes were measured within 
MIDAS\footnote{\texttt{http://www.eso.org/projects/esomidas}} through 
PSF-fitting (Stetson 1987) or aperture photometry depending on whether the 
object was point-like or extended.

\vspace{-.3cm}
\section{Results}

\begin{figure*}[t!]
\begin{center}
\hspace{.2cm}
\psfig{file=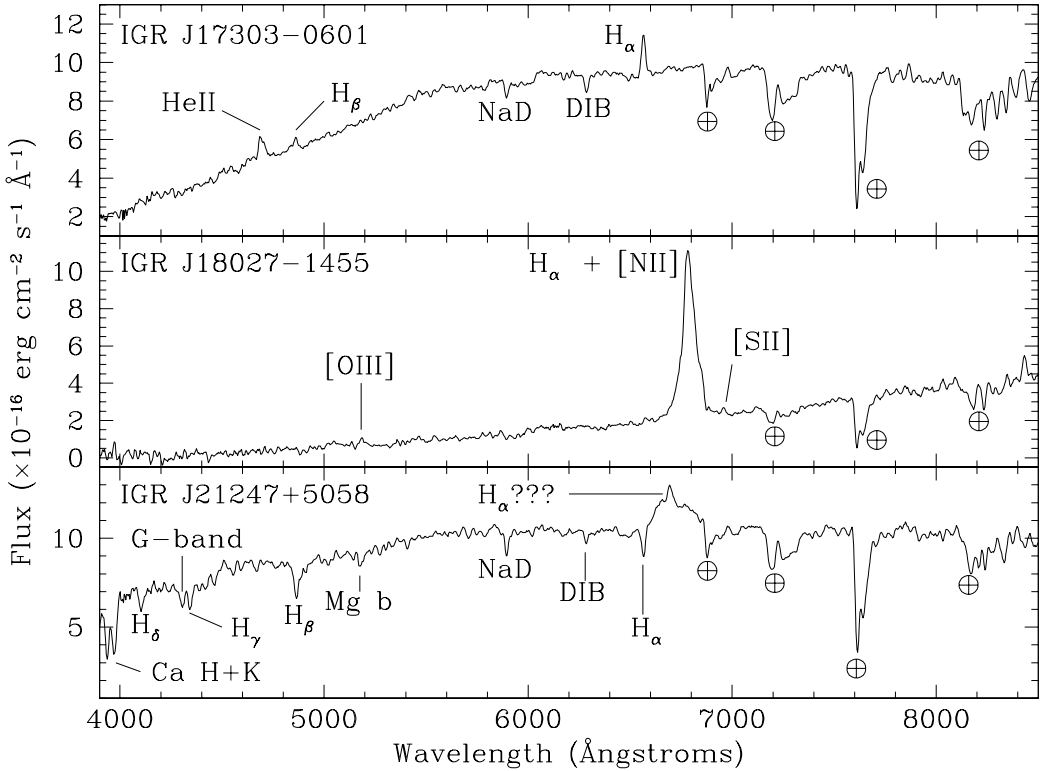,width=14.1cm,angle=270}
\end{center}
\vspace{-.6cm}
\caption{Average optical spectra of the optical counterparts to IGR 
J17303$-$0601 (upper panel), IGR J18027$-$1455 (central panel) and IGR 
J21247+5058 (lower panel) acquired with the Cassini telescope at Loiano. 
The spectra, in the 3900-8500 \AA~range, are smoothed with a Gaussian 
filter with $\sigma$=4 \AA~(i.e. comparable with the spectral dispersion).
The main spectral features are labeled. The symbol $\oplus$ indicates 
atmospheric telluric features}
\end{figure*}

\vspace{-.2cm}
In all three cases, the spectra acquired in Loiano show 
emission features which are typical of X--ray emitting objects.
We therefore conclude that we have claerly identified the optical 
counterparts to these three IGR sources. We discuss in detail the
characteristics and the nature of each case in the following.

\noindent
{\bf %The Good: 
IGR J17303$-$0601}: $R$-band imaging acquired in Loiano 
(Fig. 1, left panel) shows that the {\it ROSAT} error box, although 
small, is quite crowded, with at least 5 objects inside. The spectrum of 
the brighter object in the error box (shown in Fig. 2, top panel)
reveals the presence of the Balmer H$_\alpha$ and H$_\beta$ lines in 
emission, as well as the He~{\sc ii} $\lambda$4686 \AA~emission, 
superimposed onto a reddened continuum. Absorption lines produced by the 
Diffuse Interstellar Band (DIB) at $\lambda$6280 \AA~and by the Na Doublet 
(NaD) at $\lambda$5890 \AA~are also detected.
The spectrum of the fainter of the two proposed optical counterparts shows 
instead a lower S/N continuum on which we identified metallic 
absorption bands typical of a late-type star and no emission features. 
We therefore regard the identification of the brighter source within the 
{\it ROSAT} error circle as the optical counterpart to IGR J170303$-$0601 
as secure.

All the optical emission lines of this object are at redshift zero,
indicating that this object belongs to the Galaxy. Besides, the
presence of the He~{\sc ii} strongly indicates that this object is
undergoing mass accretion onto a compact star (e.g. van Paradijs \&
McClintock 1995). We thus conclude that this source is very likely an
X--ray binary system.

The strength of the optical Balmer emission lines can be used to estimate
the Galactic reddening toward IGR J17303$-$0601. For H$_\alpha$ and 
H$_\beta$ we measure a flux of 4.4$\times$10$^{-15}$ erg cm$^{-2}$ 
s$^{-1}$ and 9.8$\times$10$^{-16}$ erg cm$^{-2}$ s$^{-1}$, respectively.
Assuming an intrinsic Balmer decrement of H$_\alpha$/H$_\beta$ = 2.86
(Osterbrock 1989) and the extinction law of Cardelli et al. (1989), the
observed H$_\alpha$/H$_\beta$ flux ratio (i.e. 4.5) implies a 
line-of-sight reddening of $E(B-V)$ = 0.45 mag. According to Schlegel et 
al. (1998), the total Galactic color excess along the IGR J17303$-$0601 
line of sight is $E(B-V)$ = 0.61 mag: this also suggests that this source is 
within the Galaxy. We remark that this value of $E(B-V)$ is reliable 
as the source has Galactic latitude $b = +15\fdg01$ (see Schlegel et al. 
1998 for details).

$R$-band photometry returns a magnitude value $R$ = 15.78 $\pm$ 0.01 for
the object, which transforms into $R_0$ = 14.6 $\pm$ 0.1 once corrected
for interstellar extinction (we assume a 0.1 mag total error to
account for all of the uncertainties in the color excess determination).
This value is that expected from a persistent low-mass X--ray binary
(LMXB) located in the Galactic bulge: indeed, assuming a distance $d \sim
8$ kpc, we obtain an absolute magnitude $M_R \sim 0$, typical of
persistent LMXBs (van Paradijs \& McClintock 1995).
The {\it INTEGRAL} X--ray data (Bird et al. 2004) also support this 
interpretation: assuming a Crab-like spectrum, we obtain a 20--40 keV 
luminosity of 1.8$\times$10$^{35}$ erg s$^{-1}$ for a distance of 8 kpc.
This value is quite typical of persistent neutron-star LMXBs in the soft 
state (Barret et al., 2000; Masetti et al. 2004b).

\noindent
{\bf %The Bad: 
IGR J18027$-$1455}: the non-pointlike appearance of the 
optical source coincident with the radio position, along with the diffuse 
nebulosity around it (see Fig. 1, central panel) is strongly suggestive of 
an AGN as the optical counterpart to this IGR source.
Indeed, the average optical spectrum of the source (Fig. 2, central panel) 
shows a faint and reddened continuum dominated by 
a strong emission around 6800 \AA~which we readily identify with the
line complex composed of H$_\alpha$ and [N~{\sc ii}] $\lambda\lambda$6548, 
6583 at redshift $z$ = 0.035 $\pm$ 0.001. Fainter and narrower emissions
which we identify with nebular lines, that is, [O~{\sc iii}] 
$\lambda$5007 and [S~{\sc ii}] $\lambda$6716, are also found at 
wavelengths consistent with this redshift. Our photometry gives $R$ = 
16.55 $\pm$ 0.01 for the object.

The rest-frame width of the H$_\alpha$ emission, $\sim$2700 km s$^{-1}$,
allows us to classify it as a broad-line AGN, and very likely as a Type 1
Seyfert galaxy. Assuming a cosmology with $H_0$ = 65 km s$^{-1}$
Mpc$^{-1}$, $\Omega_\Lambda$ = 0.7 and $\Omega_{\rm m}$ = 0.3, this
corresponds to a distance of 166 Mpc. At this distance, the source X--ray
luminosity is 3$\times$10$^{42}$ erg s$^{-1}$ and 1.7$\times$10$^{44}$ erg
s$^{-1}$ in the 0.1--2.4 keV and in the 20--100 keV bands assuming again 
a Crab-like spectrum.
The absolute $B$-band magnitude of the object, computed assuming that 
$B$ = 19.3 (Monet et al. 2003) and considering a Galactic color excess 
$E(B-V)$ = 1.26 along the object's line of sight (Schlegel et al. 1998), 
is $M_B \sim -22$. These values place this source among the brightest 
Type 1 Seyfert galaxies detected so far (Ho et al. 1997; Malizia et al. 
1999). However, differently from the case of IGR J17303-0601, here 
the $E(B-V)$ value may not be fully reliable due to the low Galactic 
latitude ($b = +3\fdg66$) of the object.

\noindent
{\bf %The Ugly: 
IGR J21247+5058}: the spectrum of this source (Fig. 2, bottom panel) has a
puzzling appearance. It shows a smooth continuum, typical of a late F- or
early G-type star (e.g., Jaschek \& Jaschek 1987), with narrow Balmer
absorptions along with NaD, DIB, the Ca H+K doublet and the G-band and Mg
b complexes again in absorption, all at redshift 0. However, superimposed
to this stellar-like continuum, a broad emission bump around 6700 \AA~is
apparent, topped by a narrow emission. If we identify the latter as
H$_\alpha$, we obtain a redshift $z$ = 0.020 $\pm$ 0.001.

We are confident that this feature is real as it is independent of the 
position of the object along the slit, and is not detected in the spectrum 
of the (late-type, possibly K) star located 8$''$ east of the putative 
counterpart of IGR J21247+5058.

Photometry also suggest that this object is quite peculiar: the magnitude 
data collected in Loiano indicate that it has $B$ = 16.59 $\pm$ 
0.02, $V$ = 16.05 $\pm$ 0.01, $R$ = 15.46 $\pm$ 0.01, $I$ = 14.42 $\pm$ 
0.02; that is, the source seems to become substantially redded as one 
moves from the $R$ to the near-infrared. Indeed, the $BVR$ magnitudes are 
consistent with those of a late F main-sequence star located $\sim$2.5 kpc 
from Earth, while the $R-I$ color is more typical of an M-type star
(Cox 2000).

$I$-band imaging, in contrast to what is seen in other bands, seems 
to suggest that the object is slightly elongated in the east-west 
direction and with its intensity centre displaced from its PSF centroid
(Fig. 1, left panel).
We therefore put forward the hypothesis of a chance alignment between a 
Galactic F-type star and a background radio galaxy at $z$ = 0.02 which is 
responsible for both the radio morphology and the optical spectral bump 
around 6700 \AA. In this hypothesis, assuming the same cosmology as for 
IGR J18027$-$1455, the distance to this galaxy is 94 Mpc, and the 
extension of the radio lobes is $\sim$200 kpc; the 20--100 keV 
luminosity would be 1.4$\times$10$^{44}$ erg s$^{-1}$ for a Crab-like 
spectrum, locating this source also at the high end of the AGN luminosity 
distribution.

Optical and/or near-infrared imaging with sub-arcsec seeing would be 
highly desirable to confirm the alignment hypothesis presented here.
We conclude by noting that, although Mantovani et al. (1982) suggested 
that optical spectroscopy would be able to disentangle the nature of this 
object, no observations of this kind have been reported in more than 20 
years.

\vspace{-.3cm}
\section{Conclusions}

\vspace{-.2cm}
Thanks to the cross-correlation among catalogues at different wavelength, 
and by using optical spectroscopy, we established the nature of three {\it 
INTEGRAL} sources detected along the Galactic Plane: IGR J17303$-$0601 is 
a LMXB located in the Galactic Bulge, IGR J18027$-$1455 is a background 
Type 1 Seyfert galaxy at $z$ = 0.035, and IGR J21247+5058 is quite likely 
a background radio galaxy at $z$ = 0.020 with its spectrum contamined by 
the chance superposition of a Galactic star.

To conclude, we stress the fact that, in the era of large observatories,
high-quality science on up-to-date astrophysical topics, such as the hunt
for the nature of IGR sources, can still be achieved with the use of
small- and medium-sized telescopes.

\vspace{-.2cm}
\begin{acknowledgements}
We thank R. Gualandi and I. Bruni for the assistance at the Loiano
telescope. H. De Ruiter and P. Parma are thanked for useful
discussions. We acknowledge financial support by ASI (Italian Space
Agency) via contract I/R/041/02. This research has made use of the
NASA/IPAC Extragalactic Database (NED) and of the NASA/IPAC Infrared
Science Archive, which are operated by the Jet Propulsion Laboratory,
California Institute of Technology, under contract with the National
Aeronautics and Space Administration; of the SIMBAD database, operated at
CDS, Strasbourg, France; and of data obtained from the High Energy
Astrophysics Science Archive Research Center (HEASARC), provided by NASA's
GSFC. We moreover thank the anonymous referee for useful comments.
\end{acknowledgements}

\end{document}